\documentclass[twocolumn]{article}
\usepackage{graphicx}%
\usepackage{multirow}%
\usepackage{amsmath,amssymb,amsfonts}%
\usepackage{amsthm}%
\usepackage{mathrsfs}%
\usepackage[title]{appendix}%
\usepackage{xcolor}%
\usepackage{textcomp}%
\usepackage{manyfoot}%
\usepackage{booktabs}%
\usepackage{algorithm}%
\usepackage{algorithmicx}%
\usepackage{algpseudocode}%
\usepackage{listings}%
\usepackage{subfigure}
\usepackage{authblk}

\newcommand{\DE}{\mathrm{d}}

\newcommand{\VE}{\mathbf}

\newcommand{\dinteg}[4]{\int_{#1}^{#2}\!{#3}\,\DE #4}

\title{Cruciform specimens biaxial extension performance relationship to constitutive identification}
\author[1]{Gennaro Vitucci\thanks{E-mail address: gennaro.vitucci@poliba.it}}
\affil[1]{{DICATECh}, Polytechnic University of Bari, {Bari}, {Italy}}
\date{}

\begin{document}

\maketitle

\begin{abstract}
Main desired features of biaxial tests are: uniformity of stresses and strains; high strain levels in gauge areas; reliable constitutive parameters identification. Despite cruciform specimen suitability to modern tensile devices, standard testing techniques are still debated because of difficulties in matching these demands. This work aims at providing rational performance objectives and efficient cruciform specimens shapes in view of constitutive parameter fitting. Objective performance is evaluated along particular lines lying on principal directions in equibiaxial tensile tests. A rich specimen profile geometry is purposely optimized \textit{in silico} by varying cost function and material compressibility. Experimental tests, monitored via digital image correlation, are carried out for validation. New shapes are designed and tested in a biaxial tensile apparatus and show to perform better than existing ones. Parameter fitting is efficiently performed by only exploiting full field strain measurements along lines. Small gauge areas and small fillet radii cruciform specimens get closer to the ideal behavior. For constitutive parameters identification in two-dimensional tensile experiments, data analysis on gauge lines deformation suffices.
\end{abstract}

\section{Introduction}
An ideal biaxial tensile experiment would reproduce an homogeneous specimen deformation, so that predictions of constitutive models could be promptly tested on a range of deformation classes minimizing the need of \textit{ad hoc} numerical tools and inverse analysis. Nevertheless available experimental data are scarce. For instance, in the recent work \cite{linka2023new}, neural networks have been developed for rubberlike constitutive model discovery. The most recent biaxial dataset used as database in that paper is \cite{blatz1962application}, which dates back to 1962. 

Traditionally, the difficulty in obtaining reliable data have been due to lack of equipment in common research facilities. Nowadays, despite a broader availability of biaxial tensile machines, the experimental mechanics literature is still flourishing about the design of specimens which satisfy certain kinematic and static constraints upon loading since no universally accepted shapes for biaxial test exist \cite{tiernan_design_2014}. Already in the work \cite{demmerle_optimal_1993}, conceptual criteria were proposed about how an ideal cruciform specimen should behave for extracting rigorous constitutive information from biaxial tests. They can be summarized as obtaining, in a large test zone: a homogeneous deformation; a high strain level. The homogeneity is required for easily comparing predictions to experiments; the high strain level allows exploring large material deformations without incurring into failure outside the test zone. Both goals are hard to achieve, because gauge areas boundary conditions are affected by the substructures transferring loads from the device grips. 

A range of topological solutions have been proposed, based on which of the two aforementioned objectives an improvement is sought for (e.g. see reviews in \cite{demmerle_optimal_1993,avanzini2016integrated}). Clearly, boundary conditions affect strain distributions within the gauge area \cite{simon-allue_unraveling_2014}. Although linear guides, by letting the clamping zone expand without shear, have been utilized in the past (e.g. \cite{blatz1962application,kawabata1981experimental}), this system requires multiple connections, personnel interaction during tests and can not provide real time data. Clamping, on the other hand, is the standard and most gripping method, but it induces shear which affects biaxiality in the tested zone \cite{nolan_correct_2016}. As a remedy, long or slotted arms have been tested. In this way, the specimen arms result more compliant and the gauge strain level decreases \cite{zhao_novel_2014}. As an alternative, simple or double hinges and sutures or hooks have been engineered, but such discretization of load application heavily localizes strains with possible failure around holes \cite{simon-allue_unraveling_2014}. On the contrary, stiffer load transfer, obtained by reducing thickness in the test area \cite{schemmann_cruciform_2018,imtiaz_fundamental_2019,dexl_numerical_2023}, stiffening or enlarging the arms end up worsening biaxiality \cite{chen2022designing}.   

With this work, we aim at contributing toward an experimental approach which: can be both easily reproduced with common mechanical laboratory tools; tested by means widely adopted tensile machines and grips; requires minimal specimen machining for common materials usually supplied as sheets; can return useful information to be promptly compared with theoretical predictions. To this end, we deal with constant thickness, compact thin cruciform specimens, i.e. without holes or slots, whose arms are clamped in the device grips. As design variable, only the specimen profile shape is considered, by adopting a five degrees of freedom smooth single curve, in order to avoid profile discontinuity at the source of strain concentration \cite{lamkanfi2010strain}.

High biaxiality has been the goal of, among others, the shape optimizations conducted in \cite{abdelhay_newly_2009,zhu_optimal_2019,chen2022designing}. Mixed specimen quality criteria, also including high strain levels, have been adopted by \cite{makris_shape_2010,seibert_biaxial_2014,zhao_novel_2014,bauer_parametric_2016,yang_optimization_2023}. Both these performance objectives are examined also in the present work. The main novelty with respect to existing literature consists in the simultaneous adoption of: a weighted multi-objective cost function which enables investigating the extent by which either of the two objectives influences the optimized shape and performance; a systematic optimization procedure which is based on measurable, material independent errors, expressed in terms of strain invariants; the computation of errors along a line, instead that in significative points or across whole surfaces. This line evaluated error enables to establish a direct relation between immeasurable stresses and forces, by equilibrium. Error integration along lines has already been adopted by \cite{zhao_novel_2014}, but the considered lines did not statically isolate the specimen portion and the load cells from the remaining structure. As it is shown hereinafter, our choice, independent on reference frame, greatly simplifies constitutive parameters identification, without the need of computationally costly numerical simulations.

As a case study, the proposed criteria are applied, in the framework of infinitesimal strain theory, for designing cruciform samples and identifying their elastic constants.  Our optimization based design is carried out by means of the finite element method (FEM) and it prospects a set of optimal shapes depending on objectives weights and material parameters. These novel shapes are cut from a polymeric material and tested in equibiaxial experiments together with specimens borrowed from literature. The gauge area strain field is reconstructed by digital image correlation (DIC), as this is a very versatile and cost effective strain monitoring technique \cite{gower_towards_2010,ramault_comparison_2011}. The performance of different shapes is analyzed in terms of: efficiency according to the newly proposed error measures; effectiveness in parameter identification \cite{hartmann_basic_2018}.

\section{Methods}
Hereafter, the ideal deformation state of a cruciform specimen subject to a biaxial test is described, in the particular case of equibiaxiality. With this ideal goal in mind, the discrepancy between actual kinematics and performance objectives can be quantified so that a set of FEM optimized shapes can be designed. This is done by modulating a highly flexible smooth profile geometry. Subsequently, the methods for an experimental validation are developed, particularly with regard to full field strain measurements and their analysis.
\begin{figure*}[h!]
	\centering
	\includegraphics[width=.7\textwidth]{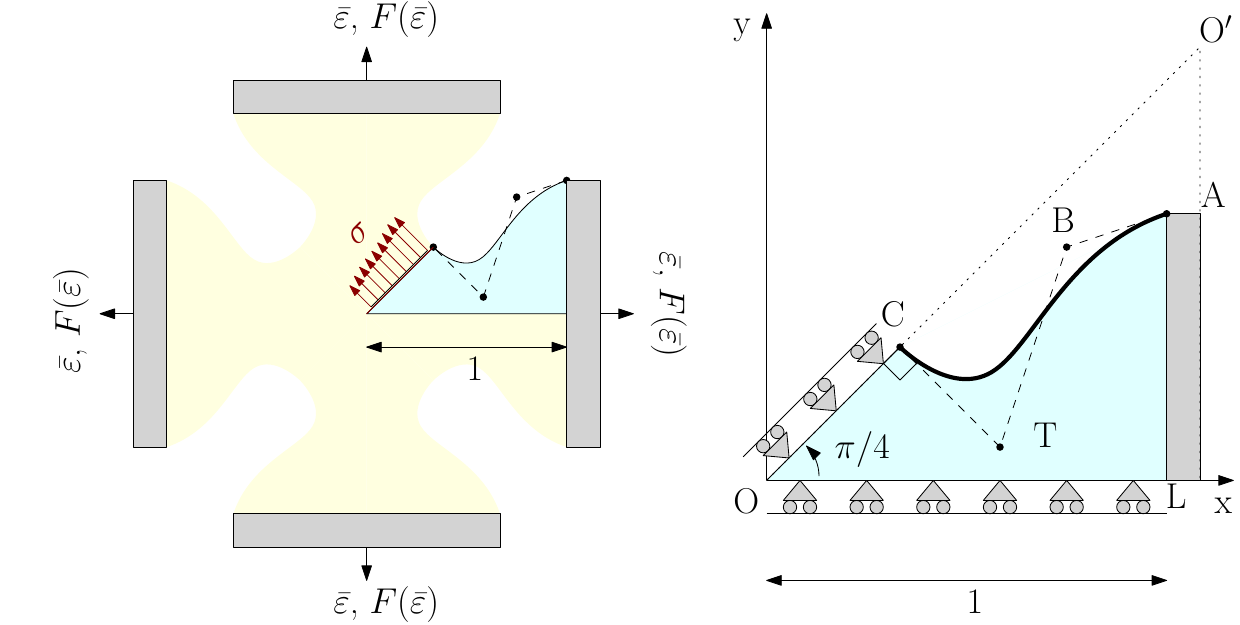}
	\caption{Geometry definition of the cruciform specimen and mechanical problem. One eighth of the specimen and its symmetry axes are represented under equibiaxial loading. Its profile is determined by cubic B-splines whose control points are A, B, T, C. The profile shape parameters are the coordinates $y_A$, $x_B$, $y_B$, $x_T$ and $x_C$. }
	\label{fig:geometry_par}
\end{figure*}

\paragraph{Performance objectives}
Perfect equibiaxiality is complicated to achieve in large areas of clamped specimens. An intuitive explanation is that an equibiaxial deformation is a homothetic transformation. Clamping, though, prevents transversal expansion of cruciform sample arms and this induces shear. In order to quantify a specimen suitability for equibiaxial tensile tests, measures of the discrepancy between an ideal and the experimental case need to be introduced. In an equibiaxial tensile test on an ideal specimen, the experimenter would impose a macroscopic stretch $\bar{\varepsilon}$ and obtain an equibiaxial deformation in a known specimen area. It is convenient to deal with principal strains, since they are independent on reference frame. Thus, in every point $\VE{X}$, the infinitesimal strain tensor, expressed in the planar principal frame, would thus take the matrix form: 
\begin{equation}
	\left[ {\begin{array}{cc}
			e_{1}(\VE{X}) & 0 \\
			0 & e_{2}(\VE{X}) \\
	\end{array} } \right]=
	\left[ {\begin{array}{cc}
			\bar{\varepsilon} & 0 \\
			0 & \bar{\varepsilon} \\
	\end{array} } \right] \forall \VE{X}\in\text{tested zone},
\end{equation}
where the subscript 1 and 2 refer to the first and second strain principal directions and $e_i$ are the principal strains, with $e_1>e_2$. As a result, once defined the maximum planar shearing strain $\Delta(\VE{X})$ and the average principal strain $\Sigma(\VE{X})$, one would aim at:
\begin{subequations}
	\begin{equation}
		\Delta(\VE{X}):=\frac{e_1(\VE{X})-e_2(\VE{X})}{2}=0.
	\end{equation}
	\begin{equation}
		\Sigma(\VE{X}):=\frac{e_1(\VE{X})+e_2(\VE{X})}{2}=\bar{\varepsilon},
	\end{equation}
\end{subequations}
Accordingly, we introduce two integral errors based on these quantities. Namely, the first one measures the loss of equibiaxiality and the second one quantifies the loss of strain as follows:
\begin{subequations}
	\begin{equation}
		\eta_\Delta=\frac{1}{\bar{\varepsilon}}\sqrt{\frac{\dinteg{\text{OC}}{}{\Delta^2(s)}{s}}{\overline{\text{OC}}}},    
	\end{equation}
	\begin{equation}
		\eta_\Sigma=\frac{1}{\bar{\varepsilon}}\sqrt{\frac{\dinteg{\text{OC}}{}{(\Sigma(s)-\bar{\varepsilon})^2}{s}}{\overline{\text{OC}}}}.
	\end{equation}
	\label{eq:err_def}
\end{subequations}

Notice that these two errors can be interpreted as standard deviations from the expected (ideal) performance and are computed on the whole gauge line OC, (see Fig.\ref{fig:geometry_par}). Whereas in previous works similar norms have already been used (e.g. \cite{hartmann_basic_2018,dexl_numerical_2023}), here it is proposed the selection of a geometric entity, for the integrals in Eq.\ref{eq:err_def}, which allows for statically isolating a part of the structure. In two dimensional settings, such an entity should be a line, since surfaces or points do not allow to directly extract resulting forces. In other words, the external force $F$ measured at the clamps level can be directly related to the average stress state $\sigma$ along the whole gauge line. This is important in view of constitutive investigations, given that, while deformations can be estimated by full field measurements, stresses can only be inferred in an integral sense from the machine load cell. Furthermore, if the strain field is known along a line, it would be computationally inexpensive to assign such a state to elements on that line and to retrieve the total exerted force instead of simulating the whole structure behavior for deducing material parameters. Other scholars have preferred obtaining uniform stress rather than strain state (e.g. \cite{chen2022designing}), but this: complicates the mathematical treatment, since constitutive equations are usually expressed in the form of stress as a function of strain; requires a device controlled in force, which is unusual.

Generally, the search for low $\eta_\Delta$ and $\eta_\Sigma$ might lead to conflicting solutions. For this reason, in view of an optimized design, we finally group them into one indicator 
\begin{equation}
	\eta=(1-\alpha)\eta_\Delta+\alpha\eta_\Sigma,
\end{equation} 
where the parameter $\alpha\in[0,1]$ is used to build a weighted sum multi-objective function. Here, a value of $\alpha=0$ or $1$ shifts all the objective towards a low $\eta_\Delta$ or low $\eta_\Sigma$, respectively.

\paragraph{Specimen shape} 
For the sake of manufacturing simplicity, we restrict our geometry to a simply connected surface, i.e. without holes, and constant thickness, as most materials are produced in sheets. Furthermore, the insertion of slots causes the arms to elongate more than the gauge area thus lowering $\eta_\Sigma$.
Cross-shaped specimens are widely used in biaxial tests because of their suitability to modern tensile machines. These, indeed, have four controllable arms where to clamp the material. Since the specimen extremities are usually bound to be rectilinear segments, the only border to be designed remains its profile. 

There could be in theory different choices for the sample profile parametrization. It has however been noted that the profile should be as continuous as possible in order to avoid stress concentrations from where cracks could nucleate (e.g. see \cite{lamkanfi_shape_2015} and literature therein). A family of functions which guarantees smoothness and flexibility is splines and in particular we opt for third degree B-splines, since B-splines are: optimal in
a mathematical sense from the point of view of support and smoothness; ubiquitous in computer-aided design which is the main tool for technical drawing  (e.g. \cite{cottrell2009isogeometric}). 	

Five parameters are selected for drawing the specimen profile, as illustrated in Fig.\ref{fig:geometry_par}, and these are the degrees of freedom of the B-spline control points A,B,T,C, where A and C have triple multiplicity. Namely, $y_A$ is the clamping length, $x_c$ individuates the coordinate of the point C along the $\pi$/4 bisector, $x_T$ controls the curvature in C, the coordinates of the point B $(x_B,y_B)$ assign the slope and curvature in A and introduce the possibility for a point of inflection along AC. Notice that the line CT is orthogonal to OC in order to preserve smoothness across the diagonal symmetry. In this way, the profile shape is defined by the vector
\begin{equation}
	\boldsymbol{\Gamma}=\left[y_A,x_C,x_T, x_B, y_B\right],
\end{equation}
with the constraint that the control points polygonal CTBA stays inscribed in the isosceles right rectangle OLO$'$ of unitary catheters.

\paragraph{Mechanical model}
The investigated flat cruciform specimen thickness is small with respect to the gauge area dimension. Since the specimen is stretched by the machine in its plane, we assume that the classical plane stress hypothesis holds. The considered boundary conditions are of Dirichlet type on the clamps, i.e. clamping imposes constant displacement in the machine arms axial direction and zero displacement in the arms transverse direction. This is representative of a typical experimental setting, where the arms are clamped inside rigidly translating grips. The displacements are imposed at the grips level, in an attempt to cause an equibiaxial state in the inner specimen area. The problem symmetries can be exploited from a computational standpoint since they allow for simulating only one eight of the area by introducing sliding constraints along the lines OC and OL (see Fig. \ref{fig:geometry_par}).

In this work the material is considered isotropic and linearly elastic. Therefore, its constitutive behavior can be fully described by two material parameters only: the Young's modulus and the Poisson's ratio $\nu$. 

\paragraph{Optimized specimen design}
Finally, in search for the most suitable specimen profile shapes, we use the shape degrees of freedom described above and formulate the following constrained optimization problem:
\begin{equation}
	\begin{aligned}
		&\min_{\boldsymbol{\Gamma}}&& \eta(\boldsymbol{\Gamma},\alpha,\boldsymbol{\mu})=(1-\alpha)\eta_\Delta(\boldsymbol{\Gamma},\boldsymbol{\mu})+\alpha\eta_\Sigma(\boldsymbol{\Gamma},\boldsymbol{\mu}),\\
		&\text{with}&&\boldsymbol{0}<\boldsymbol{\Gamma}<\boldsymbol{1},\\
		&&&x_C<x_T<2x_C,\\
		&&&x_T,y_B<x_B,
	\end{aligned}
	\label{eq:opt}
\end{equation}
where $\boldsymbol{0}$ and $\boldsymbol{1}$ are the null and all-ones vectors of the same size as $\boldsymbol{\Gamma}$. These linear inequality constraints are introduced in order to inscribe the specimen eighth in the triangle OLO$'$ as motivated in the previous paragraph. The symbol $\boldsymbol{\mu}$ collects material properties which are introduced in order to investigate the effect of constitutive features onto the specimen shape design. In this way, the solution of Eq.\eqref{eq:opt} is able to return the optimal specimen geometry as a function of the objective weights and material properties, i.e. $\boldsymbol{\Gamma}=\boldsymbol{\Gamma}(\alpha,\boldsymbol{\mu})$. 

Speaking of deformation variables only, under imposed Dirichlet conditions, the Young's modulus does not play any role since it only scales stresses and forces. On the contrary, the Poisson's ratio clearly affects strain inhomogeneity in biaxial tests, since it represents interactions between orthogonal fibres. Only its influence is consequently explored, so that in Eq.\eqref{eq:opt} $\boldsymbol{\mu}=\nu$. 

From the software viewpoint, both geometry and domain decomposition (meshing) were accomplished thanks to the open source package Gmsh \cite{geuzaine2009gmsh}. The elasticity equations were solved by using the open source FEM software Elmer \cite{malinen2013elmer} through its wrapper Pyelmer \cite{wintzer_2023_7655903}. The geometry and FEM software was interfaced via Python. In this way, we were able to conduct the multi-objective constrained optimization above. In particular, we adopted the trust region algorithm \texttt{trust-constr} available via Scipy \cite{virtanen2020scipy}. 

\paragraph{Experimental settings}
In order to validate the present approach efficacy in producing suitable shapes and a handy elastic constants identification tool, an experimental investigation on rubberlike specimens was set up. The specimens were hand cut from a 1.4mm thick EPDM sheet with a clamp-to-clamp distance of 110mm. The biaxial tensile machine, produced by Damo Srl, was controlled in displacement on four independent motors and the forces, measured by the four load cells, were monitored. A square area of side ${2}x_c$ in the specimen center was painted in white for creating a suitable speckle pattern in view of DIC analysis. The specimens arms were gripped by means of screw clamps for a depth of 15mm each. The extension velocity was set as to provide a low machine strain rate of $4\cdot10^{-4}$s$^{-1}$ for dissipating viscosity. 

The tests were filmed with a Basler acA1920-155um camera with a Sony IMX174 CMOS sensor at 2.3 MP resolution on 1920$\times$1200 pixels. The camera was positioned so that the samples center of mass appeared approximately in the frame middle. The vertical distance between lens and object plane was fixed at about 220mm. 

The obtained images were DIC analyzed by means of Ncorr, a common DIC package for Matlab \cite{blaber2015ncorr}. For the sake of simplicity in phase of data analysis, square regions of interest (ROI) were drawn over the gauge area. 

A material mechanical characterization independent on equibiaxial testing was carried out for validation purposes. A rectangular strip of 10$\times$110mm$^2$ was slowly elongated up to a 1\% strain. The DIC analysis was carried out on a square area of $10\times 10$mm$^2$ in the specimen middle. The Young modulus was estimated as 7.61MPa from the force-elongation data at the machine level. The same modulus was identified by computing the average principal strain along the sample symmetry line transversal to loading. When measured in this way, the modulus resulted 7.88MPa, i.e. the gap between the two methods was only about 3.5\%. This served as validation for the DIC analysis. Moreover, this difference is justifiable from a mechanical point of view. Clamping does not indeed produce an exactly uniform uniaxial strain on rectangular strips and therefore reading only load cells data might lead to small errors for a planar aspect ratio of 10.

Subsequently, the EPDM Poisson's ratio was inferred by DIC assisted full field strain measurement of the same uniaxial test. The average ratio of the strain parallel and perpendicular to the loading axis was measured along the symmetry line perpendicular to loading and it returned a Poisson's ratio of $\nu=0.39$. 

\paragraph{Constitutive parameters identification}
In this work the mechanical parameters identification is restricted by isotropy and small strains, so that only linear theory is used.

In plane stress isotropic elasticity, the maximum and minimum principal stresses result
\begin{equation}
	\sigma_{\text{max}/\text{min}}=\frac{E}{1-\nu^2}\left(\Sigma(1+\nu)\pm\Delta(1-\nu)\right).
	\label{eq:sigmaxmin}
\end{equation}
From Fig.\ref{fig:geometry_par}, the specimen eighth equilibrium to horizontal translation can easily be written as 
\begin{eqnarray}
	\begin{aligned}
		\frac{F}{2}&=h\sin\frac{\pi}{4}\dinteg{0}{\dfrac{x_C}{\cos\frac{\pi}{4}}}{\sigma_{\text{max}}(\Sigma,\Delta,s)}{s}\\&=h\sin\frac{\pi}{4}\langle\sigma_{\text{max}}\rangle\frac{x_C}{\cos\frac{\pi}{4}}\\
		&=h x_C \langle\sigma_{\text{max}}\rangle,
	\end{aligned}
	\label{eq:eq8th}
\end{eqnarray}
where $\sigma_{\text{max}}$ is given in Eq.\eqref{eq:sigmaxmin}, $F$ is the force read by the load cell and $h$ is the sample thickness. The symbol $\langle\cdot\rangle$ is as often adopted to represent the average along the gauge line.

If machine data only are used, no information on the distribution of strains is available except that the horizontal and vertical central lines of Fig.\ref{fig:geometry_par} are equally stretched, on average, of  $\bar{\varepsilon}$. Therefore, by taking $\Sigma=\bar{\varepsilon}$ and $\Delta=0$ along the gauge line and substituting these values into Eqs.\eqref{eq:sigmaxmin}-\eqref{eq:eq8th}, one obtains the Young modulus estimate
\begin{equation}
	\text{machine}\quad\rightarrow\quad\tilde{E}=\frac{1-\nu}{2hx_C}\frac{F}{\bar{\varepsilon}}.
	\label{eq:Emachine}
\end{equation}

A slightly more informed estimate might be produced by placing and extensometer in the specimen middle, since, as it is confirmed by DIC analysis, in that point ($s=0$) the tangential strain is null. Instead of placing such a physical device, the specimen center point DIC data was used. For doing so, $\Sigma(0)$ can directly substitute $\bar{\varepsilon}$ into Eq.\eqref{eq:Emachine} to obtain
\begin{equation}
	\text{DIC}_{(0,0)}\quad\rightarrow\quad\tilde{E}=\frac{1-\nu}{2hx_C}\frac{F}{\Sigma(0)}.
	\label{eq:dic00}
\end{equation}
On the other hand, due to small experimental strains, it is licit to apply the superposition principle to Eq.\eqref{eq:eq8th}, so that the average principal stress can be computed simply by summation of the contributions of average areal strain and average tangential strain from \eqref{eq:sigmaxmin}. This averaging leads to 
\begin{equation}
	\text{DIC}_{\Delta,\Sigma}\rightarrow\tilde{E}=\frac{1-\nu^2}{2hx_C}\frac{F}{\langle\Sigma\rangle(1+\nu)+\langle\Delta\rangle(1-\nu)}.
	\label{eq:DIC_es}
\end{equation}

\section{Results}

\begin{figure*}[h!]
	\centering
	\includegraphics[width=.6\textwidth]{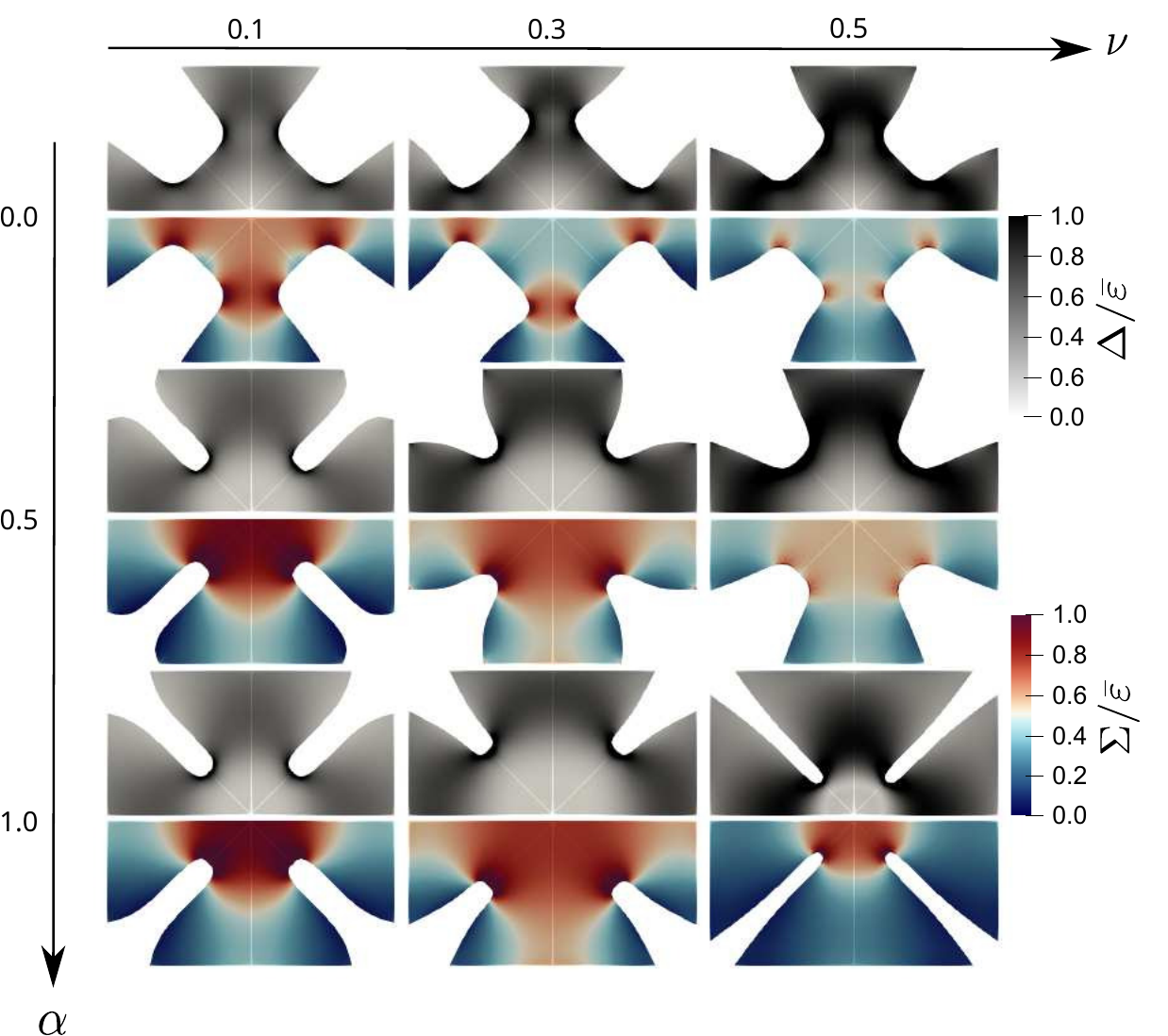}
	\caption{Effect of objective function and Poisson's ratio on the FEM optimized design and strain field. The maximum tangential strain $\Delta$ and the average principal strain $\Sigma$ are shown in the specimens upper- and lower-halves respectively with the same color mapping. The Poisson's ratio takes the value of 0.1, 0.3 and 0.5 (left to right), the objective function weight $\alpha$ is 0, 0.5 and 1 (top to bottom). }
	\label{fig:all_samples}
\end{figure*}
\begin{table*}[h!]
	\centering
	\begin{tabular}{|c c| c c| c c c c c|}
		$\nu$&$\alpha$&$\eta_\Delta$&$\eta_\Sigma$&$y_A$&$x_C$&$x_T$&$x_B$&$y_B$\\
		\hline
		0.1&	0.0	&0.31	&0.43	&0.49	&0.33	&0.54	&0.75	&0.32\\
		0.1&	0.5&	0.34&	0.07&	0.64&	0.33&	0.41&	0.81&	0.71\\
		0.1&	1.0&	0.35&	0.05&	0.70&	0.29&	0.38&	0.75&	0.64\\  \hline
		0.3&	0.0&	0.35&0.60&	0.51&	0.36&	0.64&	0.80&	0.34\\	
		0.3&	0.5&	0.45&	0.23&	0.48&	0.41&	0.47&	0.73&	0.50\\
		0.3&	1.0&	0.45&	0.18&	0.72&	0.45&	0.51&	0.67&	0.55\\  \hline
		0.5&	0.0&	0.51&	0.59&	0.44&	0.33&	0.52&	0.69&	0.32\\
		0.5&	0.5&	0.57&	0.44&	0.50&	0.38&	0.49&	0.73&	0.39\\
		0.5&	1.0&	0.71&	0.22&	0.83&	0.23&	0.27&	0.60&	0.54\\ \hline	
	\end{tabular}
	\vspace{0.2cm}
	\label{tab:opt_res}
	\caption{Optimization results. Profile shape dependency obtained via multi-objective optimization for combinations of Poisson's ratio and objective weight.}
\end{table*}
The FEM based optimization Eq.\eqref{eq:opt} was carried out by letting vary both the Poisson's ratio $\nu\in\{0.1,0.3,0.5\}$ and the objective function weight $\alpha\in\{0,0.5,1\}$ in order to explore the influence of these parameters onto the obtained shapes. The $3\times 3$ resulting profiles are shown in Fig.\ref{fig:all_samples}, where the computed maximum tangential strain $\Delta$ and half aerial strain $\Sigma$ are also depicted.

It is clear that a high $\nu$ worsens both equibiaxiality degree and transmission of strain from the clamps toward the gauge area. The simulations show indeed that the biaxiality and strain losses can greatly grow when $\nu$ increases from $0.1$ to $0.5$ (see Table \ref{tab:opt_res}). Next, optimal shapes depend on material behavior via $\nu$. Nevertheless, by keeping fixed $\alpha$, the optimal shapes do not change dramatically, as it can been inspected visually.

The multi-objective weight $\alpha$, as expected, also affects optimal strain transmission and biaxiality. High $\alpha$ appear to concentrate $\Sigma$ in the specimen center. From a geometric point of view, a high $\alpha$ requires small fillet radii and large clamping length $y_A$. This can be interpreted as follows: if the arms are large in comparison with the gauge area, they result stiffer, thus allowing for a higher deformation level in the specimen middle, similarly to reinforced arms specimens. For low $\alpha$ instead, the optimization algorithm picks large fillet radii with the effect of shifting the areas of high tangential strain far from the gauge line and towards the decentralized fillets. Quantitatively, for same $\nu$, $\eta_\Delta$ is far less adjustable than $\eta_\Sigma$.

\begin{figure}[h!]
	\begin{minipage}{\columnwidth}
		\centering
		\subfigure[Experimental setup]{\includegraphics[width=0.45\columnwidth]{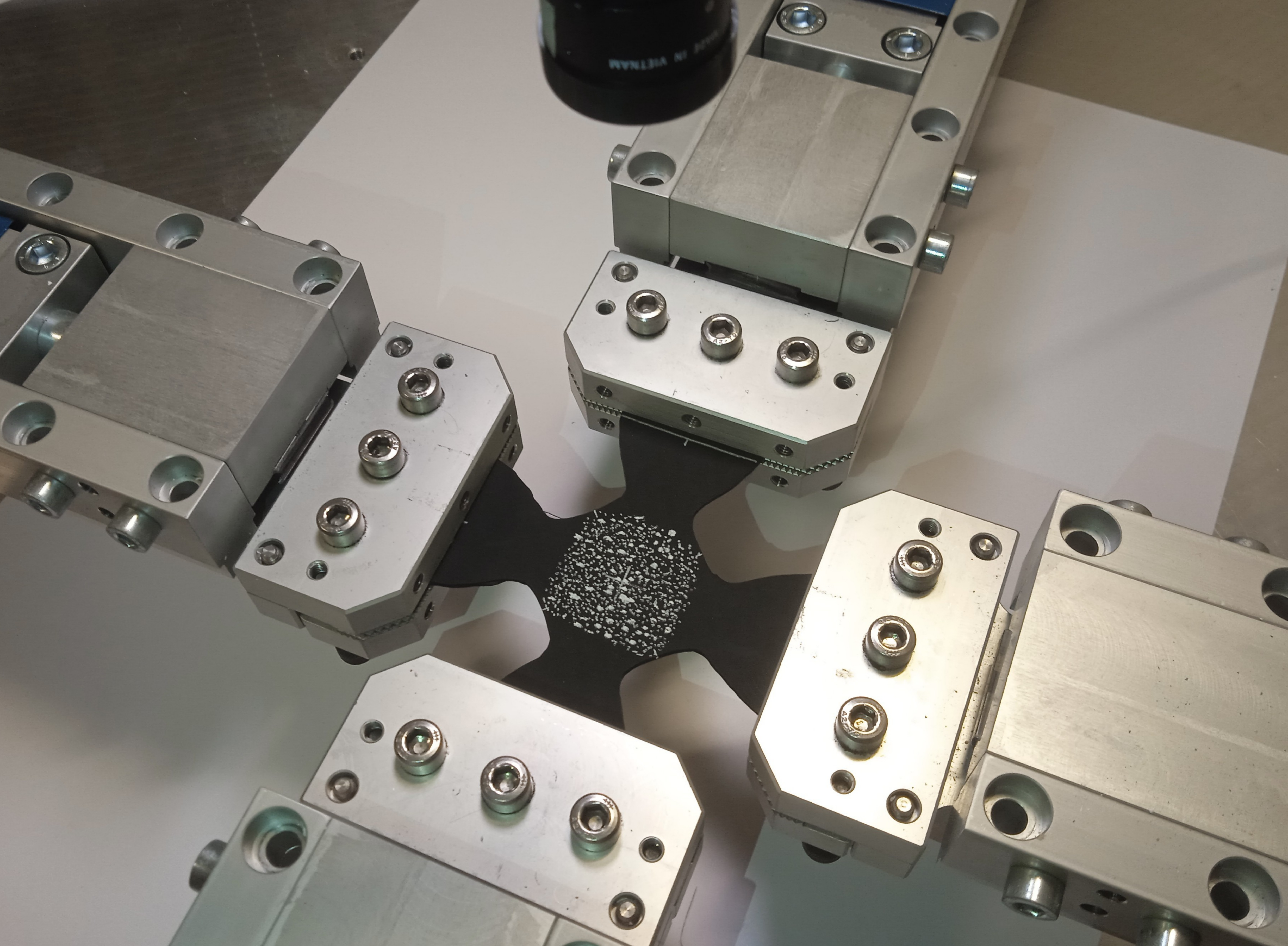}}		
		\subfigure[A00]{\includegraphics[width=0.45\columnwidth]{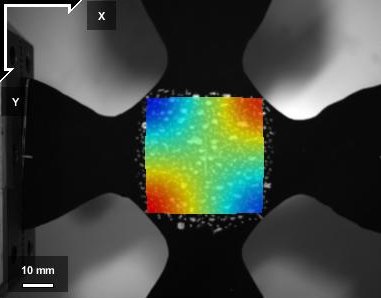}}
		\subfigure[A05]{\includegraphics[width=0.45\columnwidth]{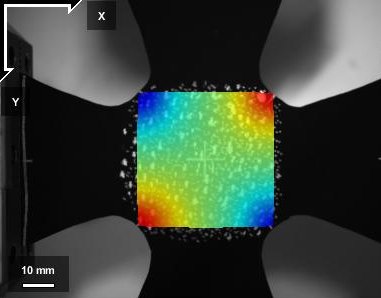}}
		\subfigure[A10]{\includegraphics[width=0.45\columnwidth]{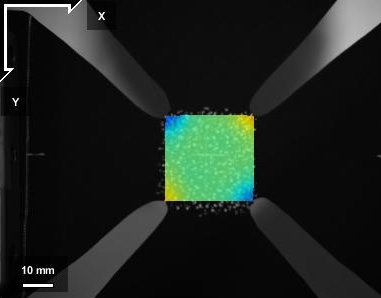}}
		\subfigure[fillet]{\includegraphics[width=0.45\columnwidth]{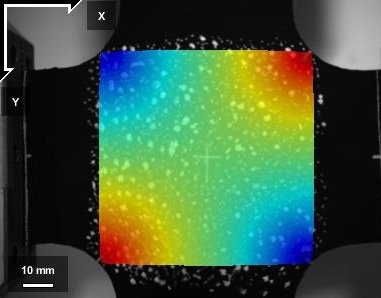}}
		\subfigure[tapered]{\includegraphics[width=0.45\columnwidth]{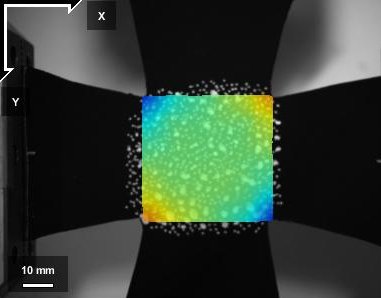}}
	\end{minipage}
	\caption{Equibiaxial tensile tests and full field measurements. Polymeric cruciform specimens were gripped in the biaxial tensile machine and subjected to equibiaxial tensile tests. The full-field strain measurements were obtained by means of DIC. In b) to f) the shearing strain $\varepsilon_{xy}$ in the Cartesian frame, as directly obtained by the DIC software, is shown. Blue and red zones, concentrated around corners, map high absolute values of $\varepsilon_{xy}$.}
	\label{fig:expset}
\end{figure}

The three optimized shapes shown in the third column of Fig.\ref{fig:all_samples}, named A00, A05, A10 hereafter, have been selected for experimental testing. The reason for this particular choice lies in an attempt to propose design solutions which can be adopted regardless of \textit{a priori} knowledge of constitutive material behavior. Since, from FEM analysis, incompressible materials result the worst performing, the choice of shapes obtained with $\nu=0.5$ appear the most conservative. Furthermore, for comparison with previously proposed designs, two more specimens, whose shapes were borrowed from \cite{avanzini2016integrated}, were cut from the same EPDM sheet. The five specimens were subjected to equibiaxial extension, in displacement control, and their gauge area full field deformation was inferred by postprocessing the filmed images in the square gauge area defined by the vertex C of Fig.\ref{fig:geometry_par}. The DIC analysis allowed returning the strains in the Cartesian frame (see Fig.\ref{fig:expset}). From these measurements, the planar strain invariants were reconstructed and our invariant based strain measures $\Delta$ and $\Sigma$ were obtained.
\begin{figure*}[h!]
	\centering	
	\includegraphics[width=.99\textwidth]{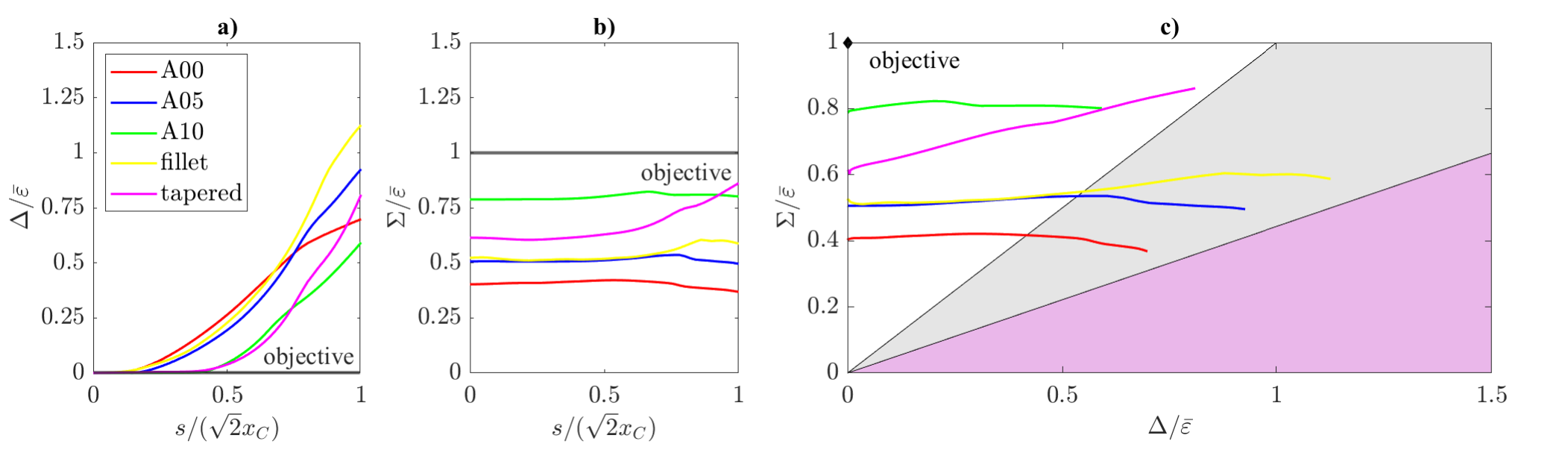}
	\caption{Experimental strains from DIC analysis. The gauge line planar strains $\Delta$ and $\Sigma$ were reconstructed from DIC analyses for the five tested specimens. All strains are normalized by the imposed machine strain $\bar{\varepsilon}=1\%$. a) and b): $\Delta$ and $\Sigma$. c) $\Sigma$ versus $\Delta$. The grey and magenta areas mean respectively contraction and compression in at least one direction.}
	\label{fig:dicsh}
\end{figure*}

\begin{table}[h!]
	\centering
	\begin{tabular}{c |c c c c c|}
		&A00&A05&A10&fillet&tapered\\
		\hline
		$\eta_\Delta$&0.39&0.42&0.23&0.50&0.29\\ 
		$\eta_\Sigma$&0.59&0.49&0.20&0.46&0.34\\ 
		\hline
	\end{tabular}
	\vspace{0.2cm}
	\label{tab:expetas}
	\caption{Experimental biaxiality and strain transmission errors.}
\end{table}
%
These ROI full field data for $\Delta$ and $\Sigma$ were interpolated along the gauge line and plotted in Fig.\ref{fig:dicsh}. In Fig.\ref{fig:dicsh}.a the five $\Delta(s)$ curves are compared. The lowest biaxiality loss were achieved by our A10 and the tapered specimens with an $\eta_\Delta$ of 0.23 and 0.29, respectively (see Table \ref{tab:expetas}). Both curves departed from the ideal zero $\Delta$ at a distance of about 50\% the gauge line length. Surprisingly, the other three specimens, i.e. A00, a05 and the fillet profiles, performed markedly worse, reaching a maximum $\Delta$ close to $\bar{\varepsilon}$ and departing from zero in the vicinity of the gauge area center of mass, even though large fillet radii resulted favorable to lower tangential strains from FEM simulations. 

Also with regards to strain transmission our A10 and the tapered specimens resulted the most efficient with a loss $\eta_\Sigma$ of 0.20 and 0.34. The latter sample was though less constant in $\Sigma(s)$, with a tendency to grow from the gauge center outwards. The other three profiles attained a higher strain loss, but all remained approximately constant along the gauge line.  
\begin{figure*}[h!]
	\centering
	\includegraphics[width=.7\textwidth]{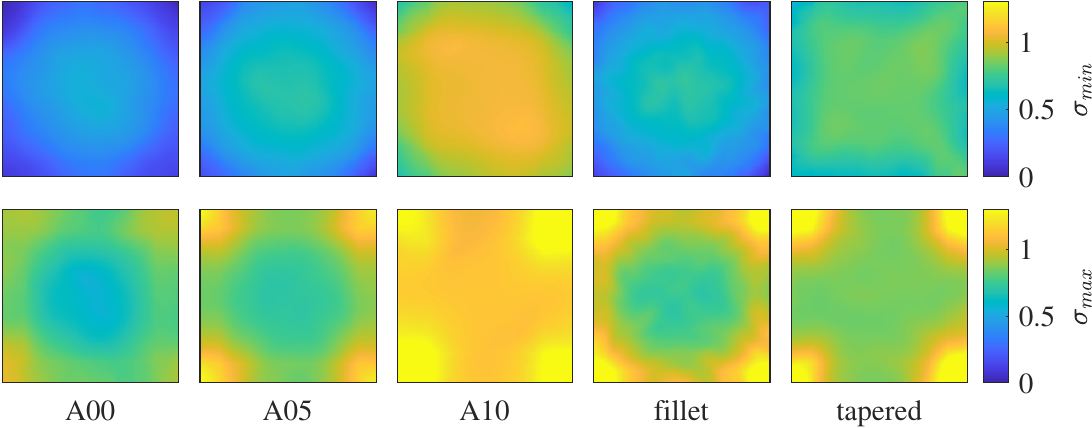}
	\caption{Minimum and maximum gauge area principal stresses. Values normalized by $E/(1-\nu^2)$.}
	\label{fig:sig_min}
\end{figure*}
In an attempt to elucidate the source of high $\eta_\Delta$ for the three specimens with large fillet radii, $\Delta$ and $\Sigma$ have been plotted against each other in Fig.\ref{fig:dicsh}.c. There, it can be observed that our A10 and the tapered specimens were the closest to the global objective of minimizing both strain and biaxiality loss.

In our tests, the principal directions of maximum and minimum normal stress are the gauge line and its perpendicular, respectively. Consequently, by adopting the minus sign in Eq.\eqref{eq:sigmaxmin}, it results that all three large fillet radii specimens reached very low positive stresses towards the fillet and a large portion of the gauge line contracted, whereas it would be expected to expand if the gauge area were uniformly stretched (grey area $\Delta<\Sigma$ in Fig.\ref{fig:dicsh}.c).

Principal stresses are mapped across the ROIs of all samples in Fig.\ref{fig:sig_min}. Very low minimum principal stresses, loosely speaking in the radial directions, were retrieved for the A00, A05 and fillet shapes, thus confirming what has been noted along the gauge line. The highest minimum gauge stress arose in A10, followed by the tapered specimen. This latter showed peculiar X-shaped stress isolines, with minimum stress loss along its ROI edges instead than corners. The maximum principal stresses, roughly the hoop ones, attained their peak around the corners and were not qualitatively different across samples. The A10 profile produced the highest mean.



\begin{figure}[h!]
	\centering
	\includegraphics[width=.99\columnwidth]{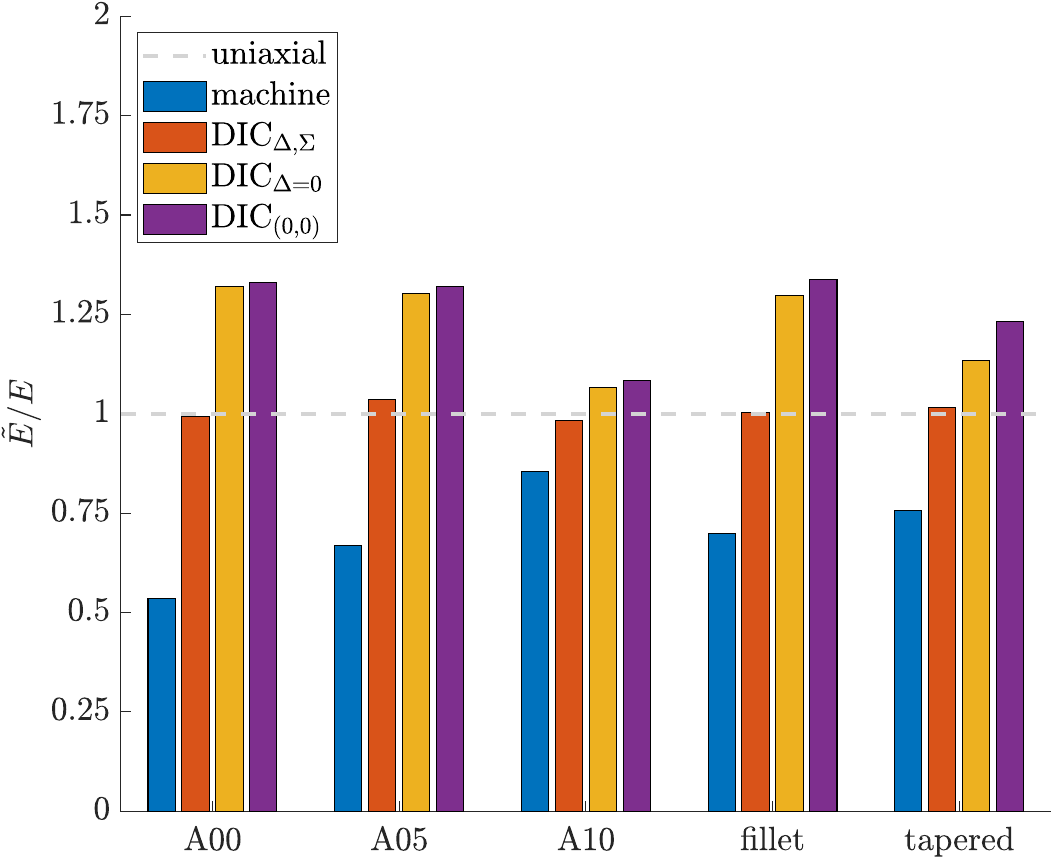}
	\caption{Parameter identification. Load cells macrostress and DIC strain measurements along the gauge line are used to infer the Young modulus already known from the uniaxial test in four different ways.}
	\label{fig:Estim}
\end{figure}

Finally, for the sake of validation and comparison, full field and load cells data were used to infer the material Young modulus, which had independently been measured in a uniaxial tensile test, as described earlier, together with Poisson's ratio. This estimates were carried out in a range of ways representing commonly adopted methods, and all of them included the knowledge of $\nu=0.39$ and the assumption of plane stress and isotropic linear elasticity. The graphical representation of these results are depicted in the histogram of Fig.\ref{fig:Estim}.

For all samples, the evaluation of Eq.\eqref{eq:Emachine} assisted only by load cells data underestimated the true value of $E$, as measured in simple extension (see Fig.\ref{fig:Estim}, blue bars). Understandably, the best so--computed Young modulus estimate corresponded to our A10 sample, since this was the closest to the \textit{blind} assumptions of perfect biaxiality and had the highest gauge strain level (see Fig.\ref{fig:dicsh}.a). Its estimate error was about 15\%. All other estimates did not pass 75\% of $E$. Again, the large fillet radii samples represented the worst case, since they exhibited both a low level of equibiaxiality and strain transmission. 
%

The evaluation procedure Eq.\eqref{eq:dic00}, simulating the use of an extensometer or biaxial strain gages in the speciment middle, differently from the machine data only exploitation, overestimated the Young modulus (see Fig.\ref{fig:Estim}, purple bars). This is expected, since $\bar{\varepsilon}/\Sigma(0)>1$ for all samples, thus the higher $\Sigma(0)$, the lower this estimate. Therefore, $\tilde{E}$ obtained form our sample A10 produced the closest guess with $\tilde{E}/E=1.08$, with the second best guess coming from the tapered profile with $\tilde{E}/E=1.23$.

Only the application of Eq.\eqref{eq:DIC_es}, for all considered specimens, allowed matching exactly the Young modulus obtained from uniaxial extension within experimental uncertainty (see Fig.\ref{fig:Estim}, red bars). Evidently, this derives from exploiting all the available DIC information about the gauge line strain field and validates the DIC readings and the linear elastic mechanical model. 

As a comparison, in Fig.\ref{fig:Estim}, yellow bars, an estimate which neglects tangential strains ($\text{DIC}_{\Delta=0}$) can be computed. In fact, by assuming perfect biaxiality within the gauge area, $\Delta=0$ can be substituted in Eq.\eqref{eq:DIC_es}. This method returns $\tilde{E}$ similar to Eq.\eqref{eq:dic00}, since $\Sigma$ is pretty constant along the gauge line ($\Sigma(0)\approx\langle\Sigma\rangle$), except that for the tapered specimen. The difference between this latter estimate and that of the full Eq.\eqref{eq:DIC_es} shows the non negligible contribution of the unwanted tangential strain $\Delta$ to the elastic problem.

\section{Discussion}
Biaxial tests are extremely useful for the investigating materials mechanical behavior. Despite a wide availability of biaxial tensile devices, the readability of experimental measurements requires care because of discrepancy between the sought deformation state and the inevitably inhomogeneous one achievable in testing. A range of approaches are usually adopted for overcoming these difficulties, but the material science community is still discussing standard testing procedures both in terms of measurement techniques and specimen shapes. 

With this work, we enrich this discussion from different points of view: performance objectives measures; newly proposed cruciform shapes; simplification of parameters identification. 

Other authors have already noticed that both biaxiality and strain transmission losses must be minimized in order to obtain easy-to-handle data and high strain levels in the gauge area. Remedies for reducing one of these two losses have repeatedly proposed, but the derived engineering solutions do not simultaneously eliminate all the shortcomings. The present multi-objective cost function includes both losses and, accordingly, establishes two metrics along a line rather than either in points or over all the sample surface. This quantification leads to simply but accurately interpret relations between stresses and measured forces, as it is demonstrated in Fig.\ref{fig:Estim}. This is crucial, since an obstacle toward wide adoption of biaxial tests has to do with the difficulty of mapping stresses to measured forces. 

The FEM simulations used for the specimen design, even though carried out by means of simplified constitutive assumptions, have allowed for a quantitative insight into the interplay between shape and behavior thanks to a flexible profile geometric parametrization. In particular, material compressibility, strain transmission and biaxiality optima lead all to different design solutions. The effect of Poisson's ratio, as expected, is to inevitably worsen the performance, based on numerical analysis. On the other hand, optimal shapes obtained with different Poisson's ratios do not differ dramatically (see Fig.\ref{fig:all_samples}). This is a relief, since it is not desirable to adopt shapes that depend sensitively on the same material parameters subject to identification, whereas an overarching experimental goal should be the adoption of standard samples for the same material class. Incidentally, this evidence goes hand in hand with that highlighted by \cite{bertin_optimization_2016} that elastic and elastoplastic specimens can be made in a similar way and still allow for correct parameter identification.

Speaking of shapes, high strain levels were achieved in our FEM based optimization by small fillet radii and large arms with a relatively small gauge area, i.e. with samples resembling tapered profiles or the one used in \cite{bell2012multiscale,putra2020biaxial}, where the optimization was although qualitative. This is reasonable, since such shapes make the arms stiffer in comparison with the gauge area. Biaxiality, instead, could in theory be improved by large fillet radii, according to our computations. Nevertheless, this improvement has failed to realize in experimental tests. ROI stress quantification confirmed extremely low  minimum stresses associated to large fillet radii, which might suggests that, compression has induced local buckling. This issue, whose detailed evaluation is out of scope here, raises an alarm in terms of experimental interpretation, since local instability can hardly be detected by force measurements or top view image processing. On the other hand, it also indicates one more reason for preferring small fillet radii and small gauge areas which have globally performed better within our metrics. In this case, indeed, all the gauge area is well conditioned in terms of tensile stresses.

Briefly, the presented experimental evidence, has shown that maximizing strain levels in the gauge area in the design process provides samples with a behavior closer to the ideal one than with shapes obtained with the goal of minimizing biaxiality losses. Besides, the effectiveness of small radii has resulted superior to the introduction of profile singularities, in the form of sharp crossroad corners, as it is demonstrated by the better behavior of the newly proposed A10 design in comparison with tapered ones.  

\section{Conclusions}
Overall, the present design procedure, based on standard deviations from ideal objectives along gauge lines which directly map average stresses to load cell forces, has allowed for the design of shapes that have performed better than existing ones, at least among simply connected surfaces, which are straightforward to create in common facilities without altering the tested materials. It has not been possible to realize an uniformly equibiaxially stretched gauge area, notwithstanding the considerable number of geometry degrees of freedoom. The objective of reading load cell data only for inferring mechanical parameters has not been reached entirely, but at best only approached by one new shape. Exclusively the use of full DIC information along the gauge line has allowed to retrieve correctly the material parameters independently measured for all samples. This surely underlines the necessity of improving the reliability and flexibility of DIC procedures in the realm of mechanical characterization \cite{reu2022dic}. 

In this work, we have exploited the fact that our gauge line lies entirely along a principal direction. This allows for testing constitutive equations which are mostly expressed in terms of strain invariants. Of course, in future studies, thinking of anisotropy or general biaxial extension, this line might not be straight and should be individuated by analyzing displacement streamlines.

These considerations help towards extensive investigations into large deformations testing and inelastic, anisotropic media, which are important frontiers, especially for soft materials, to be pushed for enlarging our knowledge of material mechanics. Even though in this study only small strains have been considered, it is encouraging that, when loading history and inelastic deformations were considered, no dramatic differences were found in optimal shapes \cite{bertin_optimization_2016}. 

As an overarching goal, standardized tests and interpretation procedures would enable the creation of material behavior databases available to scientists in many research fields. Scarcity of reliable and comparable data does indeed, at the moment, prevent proofing the experimental validity of rigorous theoretical models and relative predictions, while effective inference and computational techniques are ripe for employment \cite{linka2023new,thakolkaran2022nn,vitucci2023predictive}. 

\section*{Acknowledgments}

G. Vitucci is supported by the POR Puglia FESR-FSE project REFIN A1004.22. The author is extremely grateful to Domenico De Tommasi and Giuseppe Puglisi for fruitful discussions.

%
%
%

\bigskip
\bibliographystyle{unsrt}
\bibliography{biblio}

\end{document}